\pdfoutput=1  
\documentclass[]{article}  
\usepackage{url,float}
\usepackage{graphicx}
\usepackage{amsfonts}
\usepackage{amssymb}
\usepackage{latexsym}
\usepackage{amsmath}

\newcommand{\hide}[1]{}

\newcommand{\ABox}{
\raisebox{3pt}{\framebox[6pt]{\rule{6pt}{0pt}}}
}
\newenvironment{proof}{{\bf Proof:}}{\hfill\ABox}

\newtheorem{theorem}{{\bf Theorem}}

\newtheorem{lemma}[theorem]{Lemma}

\newtheorem{definition}[theorem]{Definition}

\newcommand{\lemlab}[1]{\label{lemma:#1}}
\newcommand{\thmlab}[1]{\label{thm:#1}}
\newcommand{\eqnlab}[1]{\label{eq:#1}}

\newcommand{\figlab}[1]{\label{fig:#1}}
\newcommand{\seclab}[1]{\label{sec:#1}}

\newcommand{\eqnref}[1]{\ref{eq:#1}}
\newcommand{\figref}[1]{\ref{fig:#1}}

\def\a{{\alpha}}

\title{From Pop-Up Cards to Coffee-Cup Caustics:\\
The Knight's Visor}
\author{%
Stephanie Jakus\thanks{
\protect\url{stephanie.jakus@gmail.com}.
}
\and 
Joseph~O'Rourke\thanks{
Dept.\ Computer Science, Smith College, Northampton,
MA 01063, USA.
\protect\url{orourke@cs.smith.edu}.
}
}

\begin{document}

\pagenumbering{roman}
\maketitle
\let\realfootnote=\footnote
\def\footnote#1{}
\tableofcontents
\let\footnote=\realfootnote

\begin{abstract}
As a pedagogical exercise, we derive the shape of a particularly elegant pop-up card design,
and show that it connects to a classically studied plane curve that is (among other
interpretations) a caustic of a circle.
\end{abstract}

\newpage
\pagenumbering{arabic}


\begin{figure}[htbp]
\centering
\includegraphics[width=0.65\linewidth]{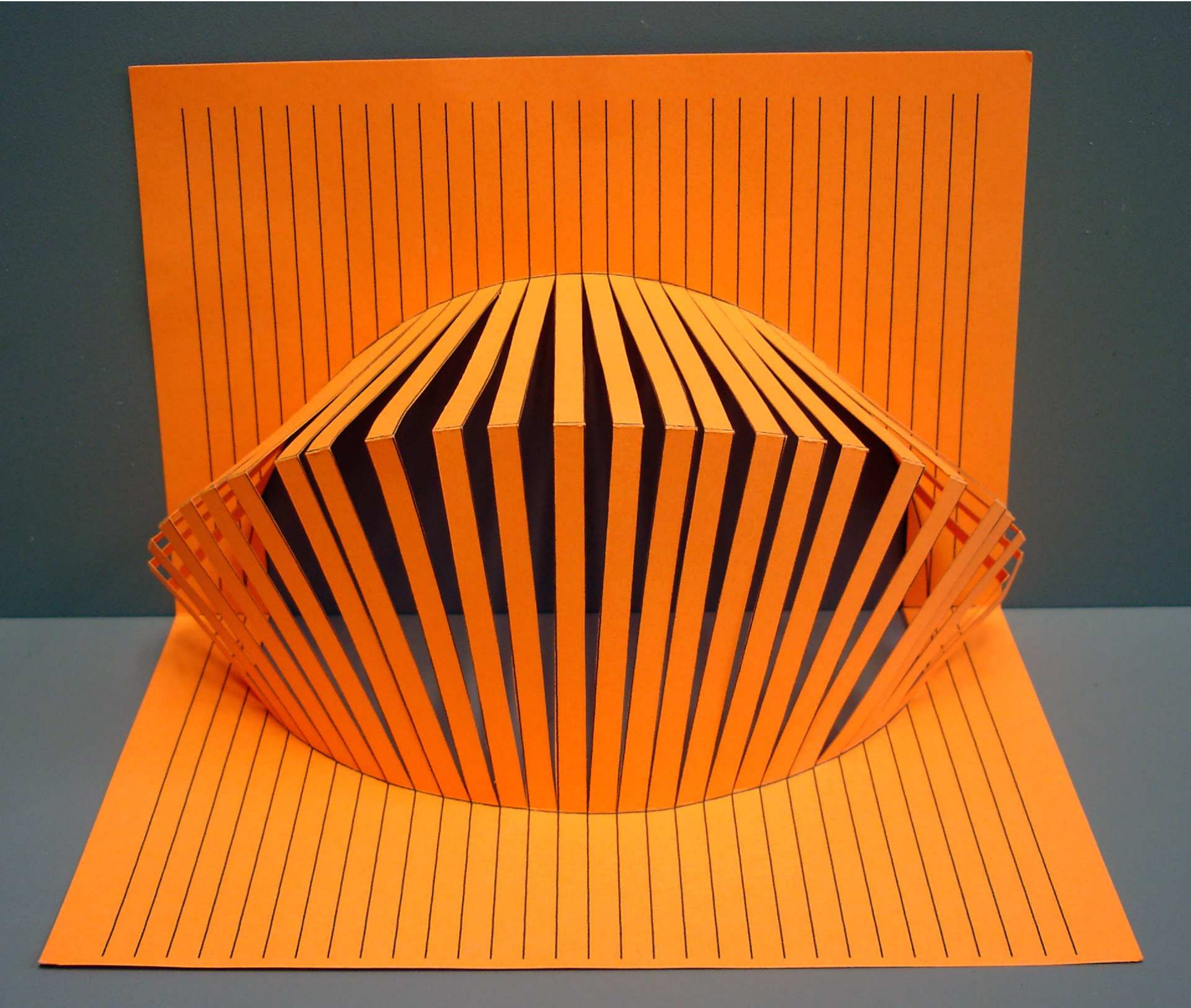}
\caption{The Knight's Visor pop-up.}
\figlab{knights.visor.photo}
\end{figure}

\section{The Knight's Visor Pop-Up}
\seclab{Knights.Visor}
There is a simple pop-up card design described  in
\cite[p.~62, Figure~III]{j-tpub-93}  as a ``multi-slit variation,''
but which we have dubbed the \emph{Knight's Visor}.
See Figure~\figref{knights.visor.photo} for the pop-up,
and
Figure~\figref{knights.visors} to justify our moniker.

The purpose of these notes is to derive equations describing the shape,
leading to some beautiful classical mathematics.
We believe this material could form an engaging module in a high-school or college geometry
or calculus course.
Given this pedagogical thrust, the derivations are carried out in some detail.
\begin{figure}[htbp]
\centering
\includegraphics[width=\linewidth]{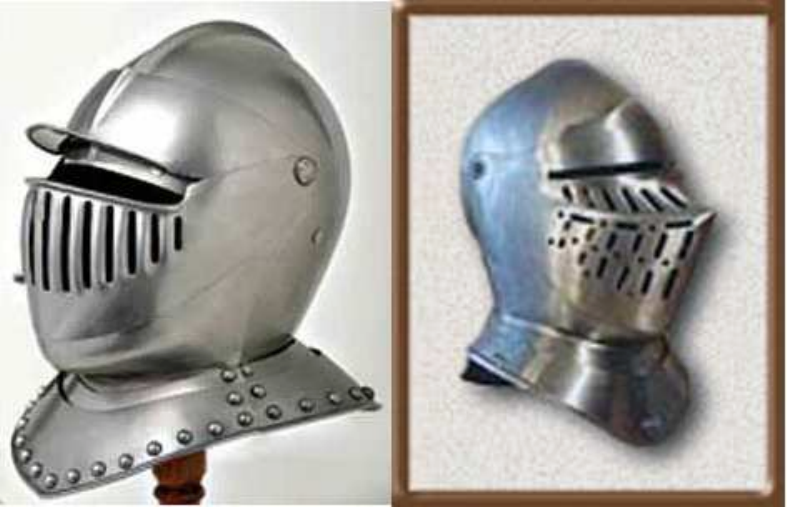}
\caption{Close helmets with visor, circa 1550.
Left:
\protect\url{http://www.by-the-sword.com/acatalog}
Right: 
\protect\url{http://www.gemstone.play.net/etimes}
}
\figlab{knights.visors}
\end{figure}

\begin{figure}[htbp]
\centering
\includegraphics[width=0.8\linewidth]{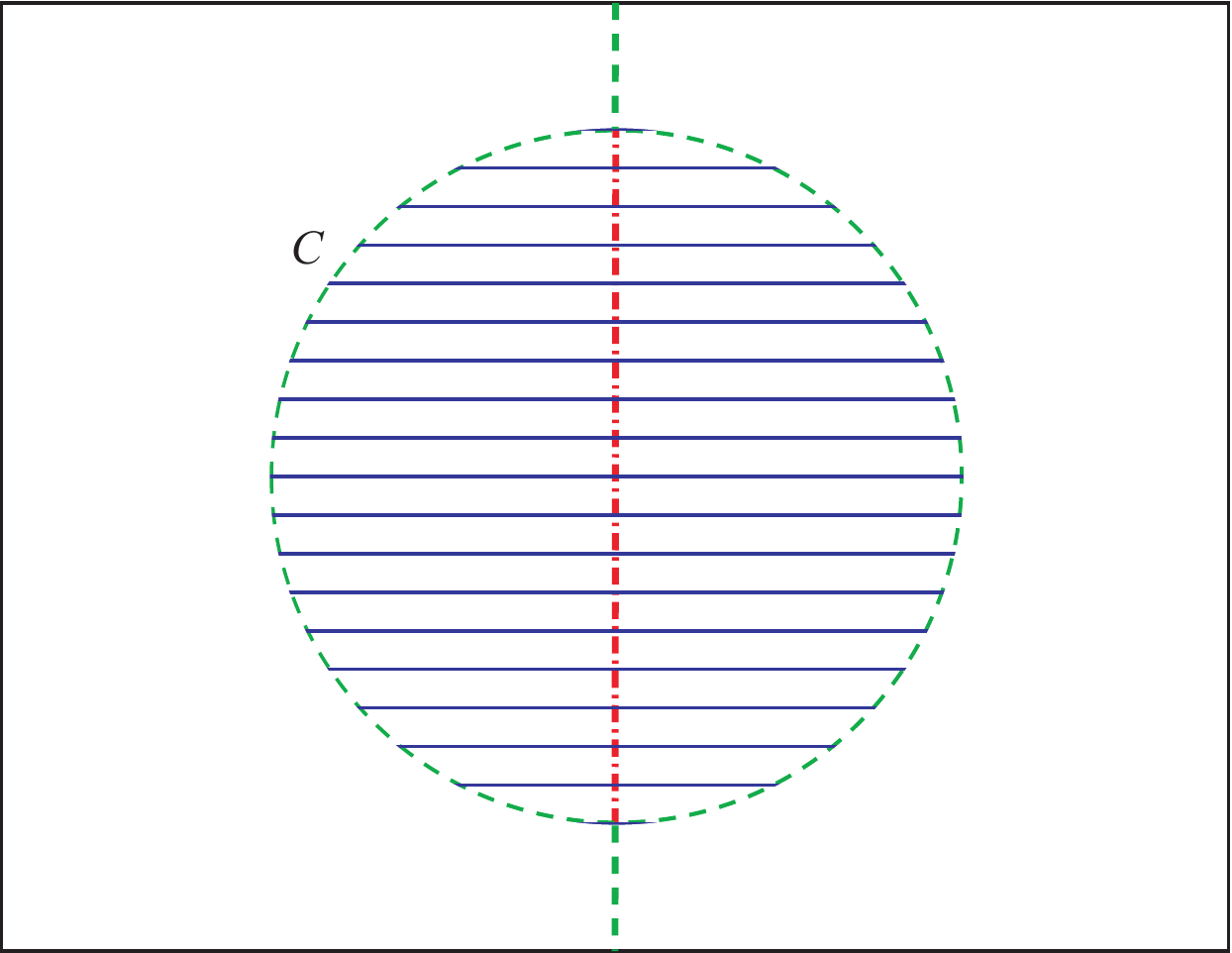}
\caption{Template for the Knight's Visor.}
\figlab{knights.visor.template}
\end{figure}

The design is remarkably simple.
Draw a circle $C$ centered on the card's centerline, and cut equally-spaced,
parallel cuts terminating on the circle boundary.
See Figure~\figref{knights.visor.template}.
This is most easily accomplished by first folding the card in half, and cutting with scissors from the centerline
to the semicircles on the front and back of the card
(see Figure~\figref{card.notation}(a) for notation.)
The creases across the diameter of $C$ are mountain creases.
Each strip ends on the circle, where it joins there in a valley crease.
These valley creases are straight segments that approximate the circle;
the creases are \underline{not} parallel to the card centerline.

\begin{figure}[htbp]
\centering
\includegraphics[width=0.75\linewidth]{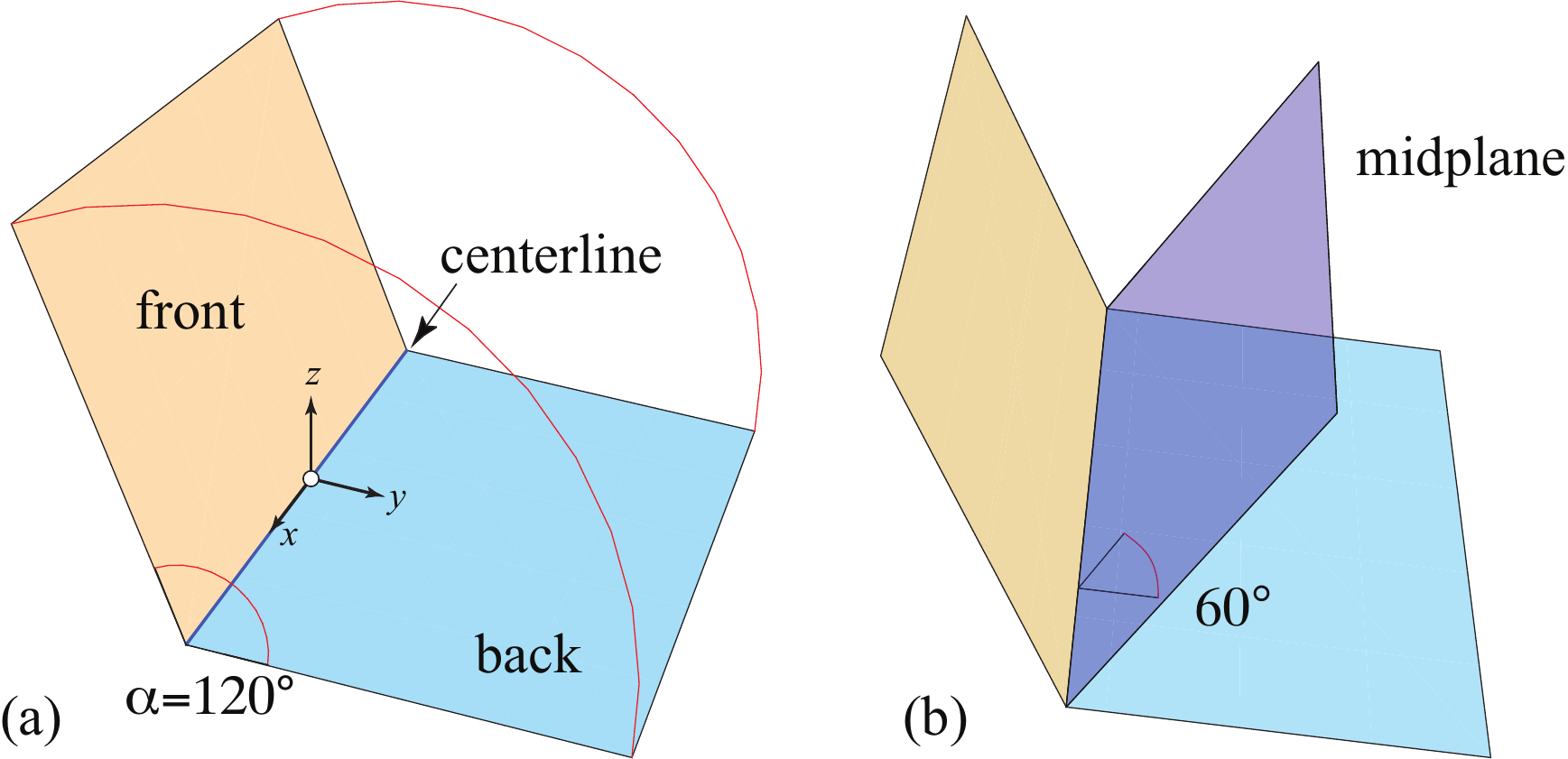}
\caption{(a)~Basic card notation.  
The dihedral angle between the card front and back is $\a$.  The card back serves
as the $xy$-plane.
(b)~The midplane is at angle $\frac{1}{2}\a$.}
\figlab{card.notation}
\end{figure}

As the card opens and closes, the motion is quite intricate.
As we will see in more detail later,
each half-strip (a ``rib") attached to the back
of the card rides on a cone (see ahead to Figure~\figref{AllCurvesCone})
whose axis is along its angled crease, 
and the half-strip rib attached to the front face of the card rides on another cone,
but this time attached to the front face.  And all the rib cone axes make different
angles with the centerline.  The combination produces a bowing that is small for long strips
and gradually increases for the shorter strips.
The result is a pleasant surprise as the card gradually closes from
fully opened (card angle $\a=\pi$), a shape variation difficult to predict from 
its mundane start in Figure~\figref{knights.visor.template}.
With thin ribs carefully cut and creased in quality card stock, 
the visor shape is almost a work of art with the angle fixed to $\a=\pi/2$.

\section{Flat Visor Curve}
Call the midcurve of the knight's visor its \emph{rim}.
It originates from the portion of the centerline of the card within the disk $D$
that has been cut with parallel slits into \emph{ribs}, each rib $R$ a thin
trapezoid between the card centerline and the boundary circle $C = \partial D$ of the disk.
We know the shape of the rim when the card is fully opened ($\a=\pi$): it is a diameter
of $C$.
Our first goal is to derive the shape of the rim when the card is fully closed ($\a=0$), which
we call the \emph{flat visor curve}, ``flat" because the visor is flattened with the card closure.

Because the valley crease where a rib meets $C$ follows the contour of $C$, and is a straight segment crease,
that crease is tangent to $C$. 
When the card is fully opened, each rib is perpendicular to the diameter of $C$ along the card centerline.
When the card is fully closed, each rib has rotated to lie completely external to the disk $D$.
See the yellow region in Figure~\figref{visor.2D}.

Now we establish a coordinate system and labels for various points to allow us to
derive an analytical expression for the equation of the rim on the card back when $\a=0$.
Let $C$ have unit radius, and be centered on the origin, with the $x$-axis through
the card centerline (see Figure~\figref{card.notation}).  We will reduce each rib $R$ to a segment, $R=ab$, with
$a=(s,0)$.
We seek the point on the flat visor curve corresponding to $a$, a function of the parameter $s$.
$C$ can be described by the equation $y=\sqrt{1-x^2}$, and so
$b=(s, \sqrt{1-s^2})$. 
Let 
$$
r = |R| = \sqrt{1-s^2} \; ;
\eqnlab{req}
$$
so $b=(s,r)$.
Next we find the tangent to $C$ at $b$, and use that to determine the \emph{reflected rib} $R'$,
reflected because the crease of the rib $R$ is tangent to $C$ where it attaches at $b$.
Refer throughout to Figure~\figref{visor.2D}.


We can find the slope of the tangent line to $C$
at $b$ in two ways: either computing $\frac{dy}{dx}$ from $y=\sqrt{1-x^2}$,
or using the fact that the tangent is perpendicular to the radial segment from
the origin to $b$.  Either way we see that the slope of the tangent at $b$
is $-\frac{s}{r}$, and from this we derive an equation for the tangent line:
\begin{equation}
y = (s-x) \frac{s}{r} + r \; .
\eqnlab{tangent.line}
\end{equation}
Now we compute the reflected rib $R'$, by computing the reflection $a'$ of $a$ across this tangent line.

\begin{figure}[htbp]
\centering
\includegraphics[width=0.95\linewidth]{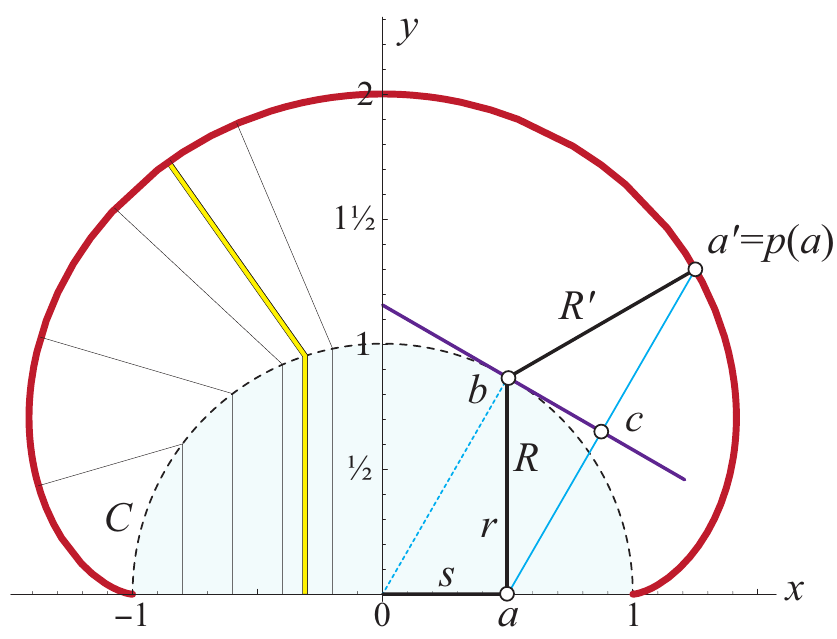}
\caption{The yellow region shows two positions of a rib: 
card fully opened ($\a=\pi$) and fully closed ($\a=0$).
The rib $R=ab$ reflects across the tangent at $b$ to $R'=ba'$.
}
\figlab{visor.2D}
\end{figure}

The segment $a a'$ is perpendicular to the tangent, and so has slope $\frac{r}{s}$,
and therefore the line containing that segment has equation
\begin{equation}
y = (x-s) \frac{r}{s} \; .
\eqnlab{perp.line}
\end{equation}
Intersecting this line with the tangent line by solving
Eqs.~\eqnref{tangent.line} and~\eqnref{perp.line} simultaneously
yields coordinates for the midpoint $c = (c_x,c_y)$ of $a a'$:
\begin{eqnarray*}
c_x & = & \frac{s( s^2 +2 r^2)}{s^2+r^2}  \\
c_y & =  & \frac{r^3}{s^2+r^2}  \; .
\end{eqnarray*}
Using $s^2 + r^2 = 1$ (because $C$ has unit radius)
and the definition of $r$ (Eq.~\eqnref{req}) reduces this to
\begin{eqnarray}
c_x & = &   s (2 -s^2) \nonumber \\
c_y & =  &  (1-s^2)^{3/2} \; .
\eqnlab{midpoint}
\end{eqnarray}
So now we can compute $a'= (a'_x,a'_y)$ from $c= \frac{1}{2}(a + a')$:
$a' = 2(c - a) + a$: 
\begin{eqnarray}
a'_x =x(s) &=& s (3-2 s^2) \nonumber \\
a'_y =y(s) &=& 2(1-s^2)^{3/2} \; .
\eqnlab{aprime}
\end{eqnarray}
We have now arrived at parametric equations for the flat visor curve.
Each $s \in [-1,1]$ leads to a point $p(s)  =  (x(s),y(s))$ on the curve.
Note that for $s=\pm 1$, $p(s) = (\pm1,0)$, and the curve is
tangent to the $x$-axis there.
For $s=0$, $p(s) = (0,2)$,
 corresponding to the centermost rib of length $r=1$ reflecting across the horizontal
tangent at the top of the circle $C$.

\section{Implicit Equation}
We make as our next goal to obtain an implicit form of the equation of the curve.
First, rewrite the parametric equations (Eq.~\eqnref{aprime}) in terms of $r$ via Eq.~\eqnref{req}: 
\begin{eqnarray*}
x &=& 2 s r + s \\
y &=& 2r^3 \; .
\end{eqnarray*}
Solving the second for $r$ yields $r = (y/2)^{1/3}$.
Substituting that into the first gives
$$
x = s + 2^{2/3} s y^{1/3} \; ,
$$
and solving this for $s$:
$$
s = \frac{x}{1+ 2^{2/3} y^{1/3}} \;.
\eqnlab{seq}
$$
Recall that $s^2 + r^2 = 1$. 
Substituting Eq.~\eqnref{seq} for $s$ and Eq.~\eqnref{req} for $r$ 
into this circle equation yields:
\begin{eqnarray*}
\frac{x^2}{\left({2}^{1/3}
   y^{2/3}+1\right)^2}+\frac{y
   ^{2/3}}{2^{2/3}} &=& 1 \\
2^{2/3} x^2+\left({2}^{1/3}  y^{2/3}+1\right)^2
   y^{2/3} & = & 2^{2/3}
   \left({2}^{1/3}
y^{2/3}+1\right)^2 \\
   2^{2/3} \left(x^2+y^2\right) & = & 3
   y^{2/3}+2^{2/3} \; ,
\end{eqnarray*}
and finally:
\begin{eqnarray}
x^2+y^2 & = & 3 (y/2)^{2/3} + 1  \eqnlab{two-thirds}\\
(x^2+y^2-1)^3 & = & \frac{27}{4} y^2 \;.
\eqnlab{implicit}
\end{eqnarray}

\section{The Nephroid}
This last form of the implicit equation, Eq.~\eqnref{implicit},
reveals it to be a
\emph{nephroid},\footnote{
I thank Sylvain Bonnot and Francesco Polizzi for identifying 
the form of Eq.~\eqnref{two-thirds} as a nephroid.
See \url{http://en.wikipedia.org/wiki/Nephroid},
\url{http://mathworld.wolfram.com/Nephroid.html}, or
\url{http://xahlee.org/SpecialPlaneCurves_dir/Nephroid_dir/nephroid.html}.
}
a planar curve studied since the 17th century.
A nephroid
is usually represented the form, 
$$
 (x^2+y^2-4t^2)^3=108t^4y^2 \;,
$$
with a parameter $t$.
For $t=\frac{1}{2}$, $4 t^2 = 1$ and $108 t^4 = \frac{27}{4}$,
and we reach exactly
Eq.~\eqnref{implicit}.
The name of this curve derives from the Greek
\emph{nephros}, which means ``kidney,"
for when the curve is drawn to both sides of the $x$-axis
(or on both the card front and card back), it is indeed
kidney-shaped:
see Figure~\figref{Kidney}.
\begin{figure}[htbp]
\centering
\includegraphics[width=0.5\linewidth]{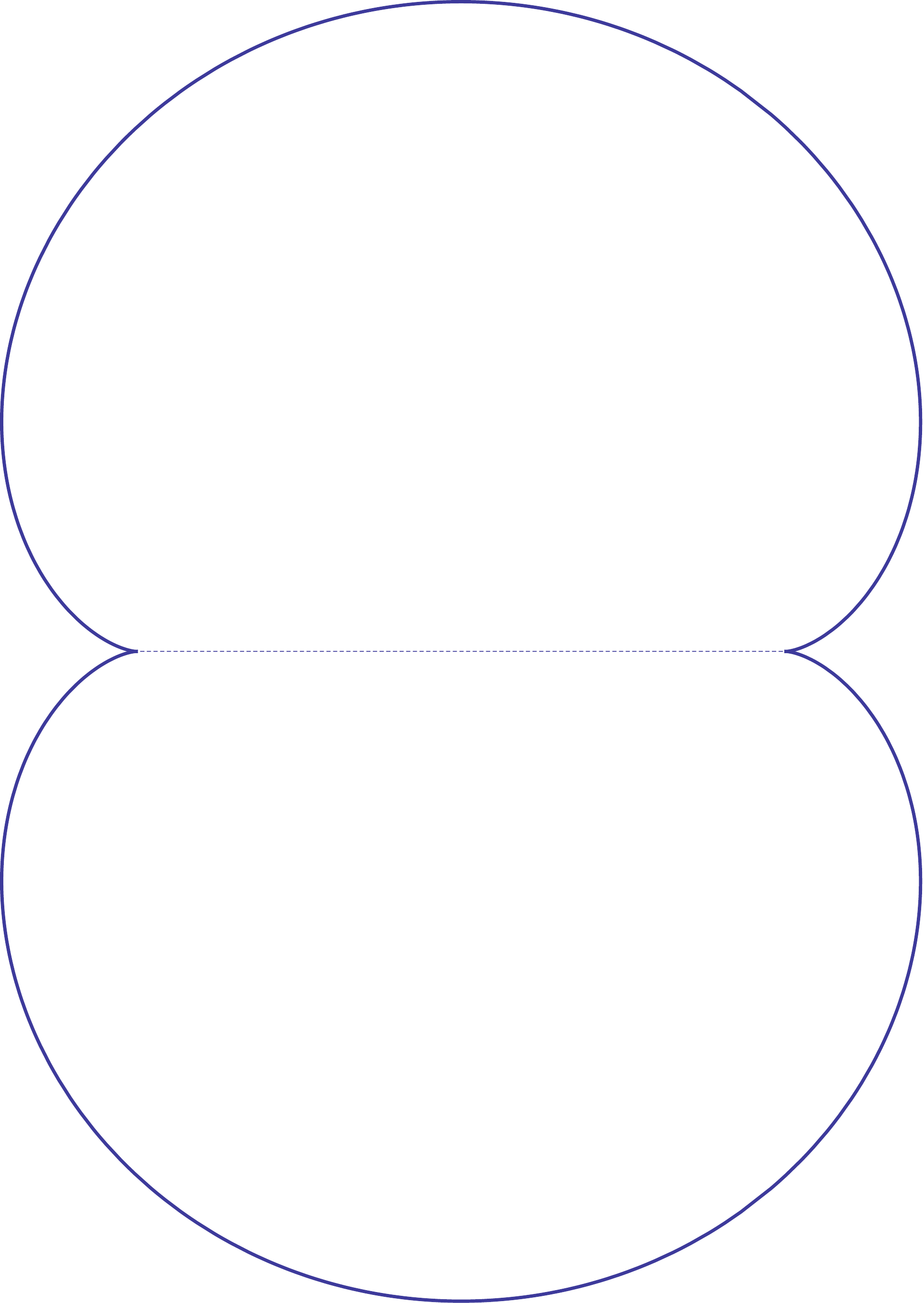}
\caption{Reflecting the visor curve over the $x$-axis.}
\figlab{Kidney}
\end{figure}

The nephroid is a rich curve with several interpretations.
For example, it is the trace of a point on one circle as it rolls on another;
see Figure~\figref{Nephroid1}.
\begin{figure}[htbp]
\centering
\includegraphics[width=0.85\linewidth]{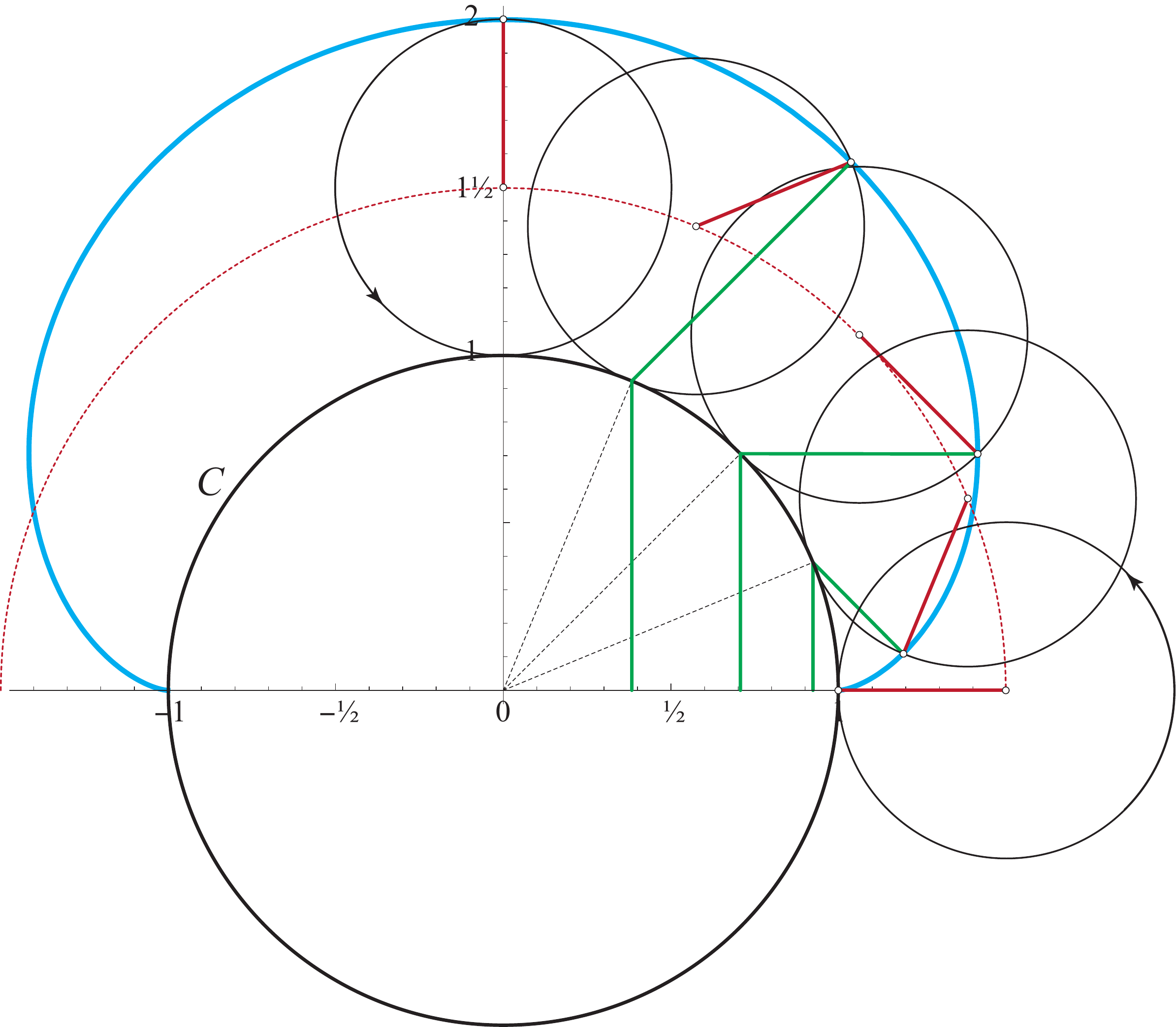}
\caption{The circle of radius $\frac{1}{2}$ rotates without slippage on $C$.}
\figlab{Nephroid1}
\end{figure}

The interpretation most natural in our context is that our nephroid is
the envelope of circles centered on 
$C$ and tangent to the diameter of $C$ on the $x$-axis.
Figure~\figref{Nephroid2} shows that two radii of these circles are the two ribs $R$ and $R'$:
$R$ perpendicular to the $x$-axis, and $R'$ perpendicular to the visor curve, where
the circle is tangent to the curve.
\begin{figure}[htbp]
\centering
\includegraphics[width=0.85\linewidth]{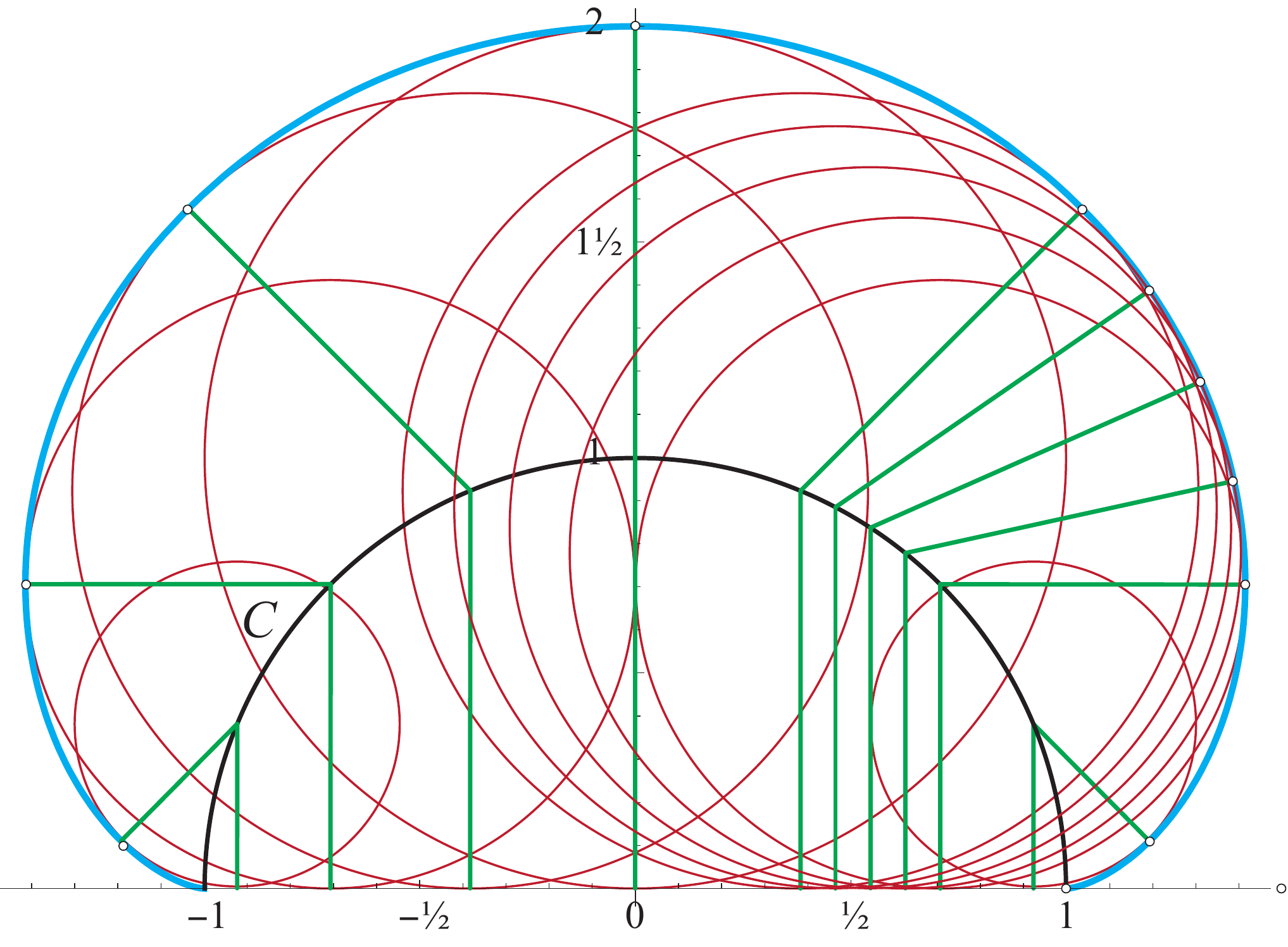}
\caption{The flat visor curve is the envelop of circles centered on $C$ and tangent to
the $x$-axis. Several rib-radii are shown (in green).}
\figlab{Nephroid2}
\end{figure}

We mention one more interpretation of the nephroid.  If the ribs $R'$ in
Figure~\figref{Nephroid2} are reflected inside $C$ rather than reflected outside,
then we can view the collection of ribs $R$ as parallel light rays reflecting
off a interiorly mirrored circle $C$.  The envelope of these reflected rays is
a rotated nephroid of half the size, as illustrated in
Figure~\figref{Nephroid3}.  
We do not pause to derive this connection~\cite[p.~158]{w-pdcig-91}, but
only marvel at it.
\begin{figure}[htbp]
\begin{minipage}[h]{\linewidth}
\centering
\includegraphics[width=0.85\linewidth]{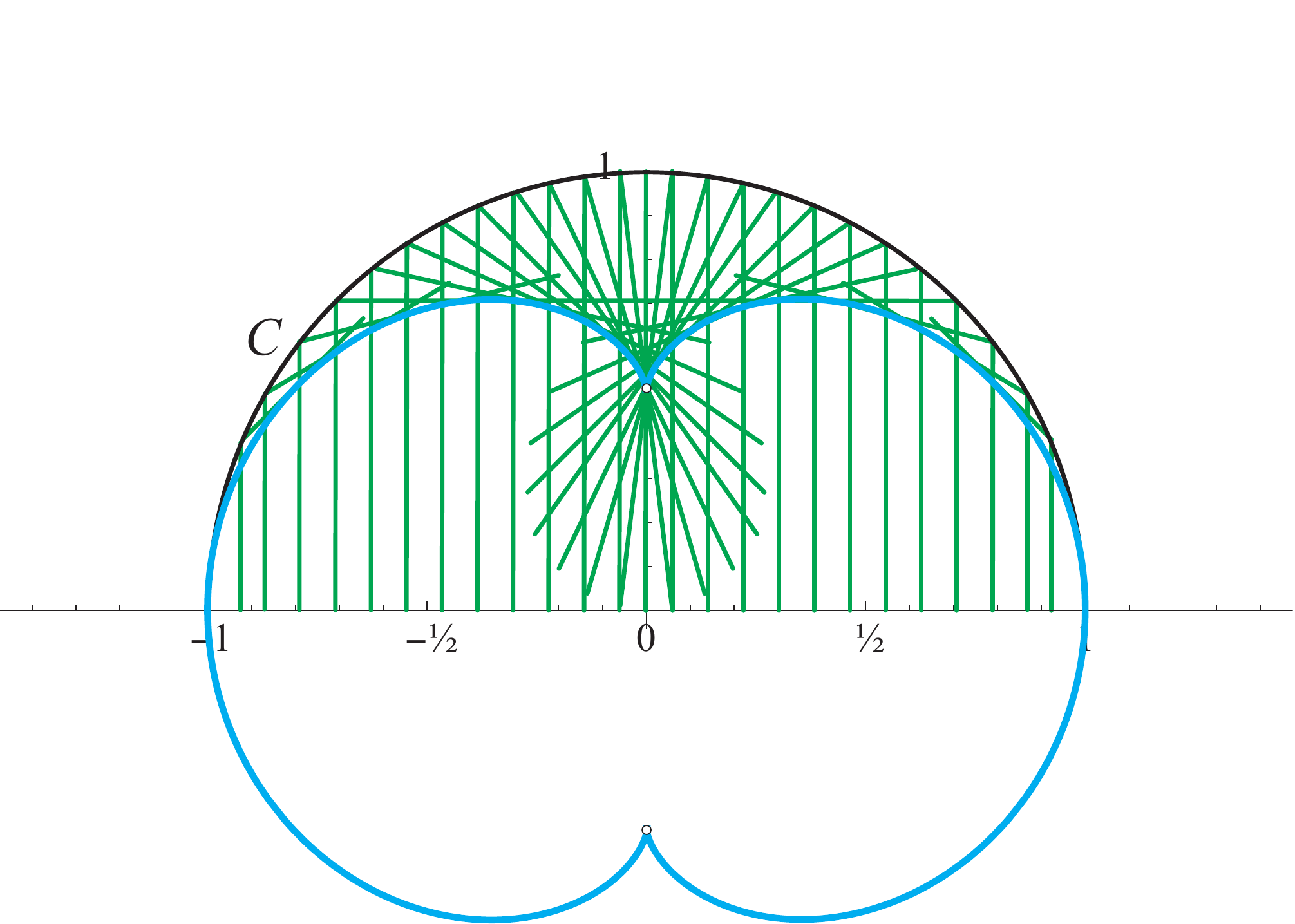}
\caption{Caustic formed by parallel light rays reflecting inside $C$.}
\figlab{Nephroid3}
\vspace{24pt}
\centering
\includegraphics[width=0.5\linewidth]{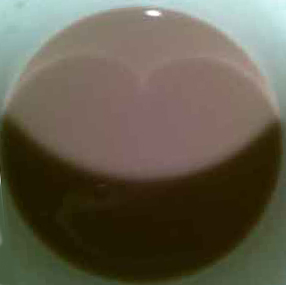}
\caption{Caustic in glass coffee mug.  
[Wikimedia Commons:
\protect\url{http://en.wikipedia.org/wiki/File:CardioidInCoffeeGlass.jpg}]
}
\figlab{CardioidInCoffeeGlass}
\end{minipage}
\end{figure}
The concentration of crossing reflected lightrays forms a ``caustic'' (or ``catacaustic")
which---remarkably---can be seen
in a drinking glass or a coffee mug
under the right circumstances,
as demonstrated in
Figure~\figref{CardioidInCoffeeGlass}.

\section{Visor Curve in 3D}
Next we derive an equation for the visor curve in 3D.
Recall that the dihedral angle between the card front and back is $\a$.
We seek parametric equations for points on the curve $p(\a,s) = (x,y,z)$
as a function of both $\a$ and the same parameter $s$ we used for the flat visor curve.

It is clear that the visor curve lies in the medial plane $M$ (see Figure~\figref{card.notation}(b)), which makes
an angle $\tfrac{1}{2} \a$ with the card back. 
Thus
$$
z(\a,s) = \tan (\tfrac{1}{2} \a) \;  y(\a,a) \; .
\eqnlab{zy}
$$
We consider the card back, in the $xy$-plane, fixed while the card front rotates.
Now imagine fixing $s$ and so fixing $a=(s,0)$ and $b$.
Because $p=p(\a,s)$ is connected to point $b$ by the rib $R$,
$p$ is a distance $r=|R|$ from $b$, and so it lies on a sphere $S$ of that
radius centered on $b$.
Let $b_1$ be the analog of point $b$ but on the card front instead of the card back.
Then again $p$ lies a distance $r$ from $b_1$, on sphere $S_1$,
which moves as $\a$ changes.
The spheres $S$ and $S_1$ intersect in a circle that lies in $M$, a circle that projects to the $xy$-plane
as an ellipse.
We can represent the two-spheres constraint as
\begin{eqnarray}
(b - p) \cdot (b - p) &=& r^2 = 1-s^2 \\
(b_1 - p) \cdot (b_1 - p) &=& r^2 = 1-s^2 \; ,
\eqnlab{spheres}
\end{eqnarray}
where $b_1 = (s, r \cos \a, r \sin \a)$.

One more constraint will pin down $p(\a,s)$.
Returning to Figure~\figref{visor.2D}, note that the rib in 3D at all times forms the same
angle with the tangent to $C$ at $b$, because the crease at the base of the rib lies along that tangent.
Thus as the rib rotates in 3D, attached all the while to $b$, it sweeps out a cone
whose rim is a semicircle perpendicular to the $xy$-plane.
See Figure~\figref{AllCurvesCone}.
\begin{figure}[htbp]
\centering
\includegraphics[width=0.85\linewidth]{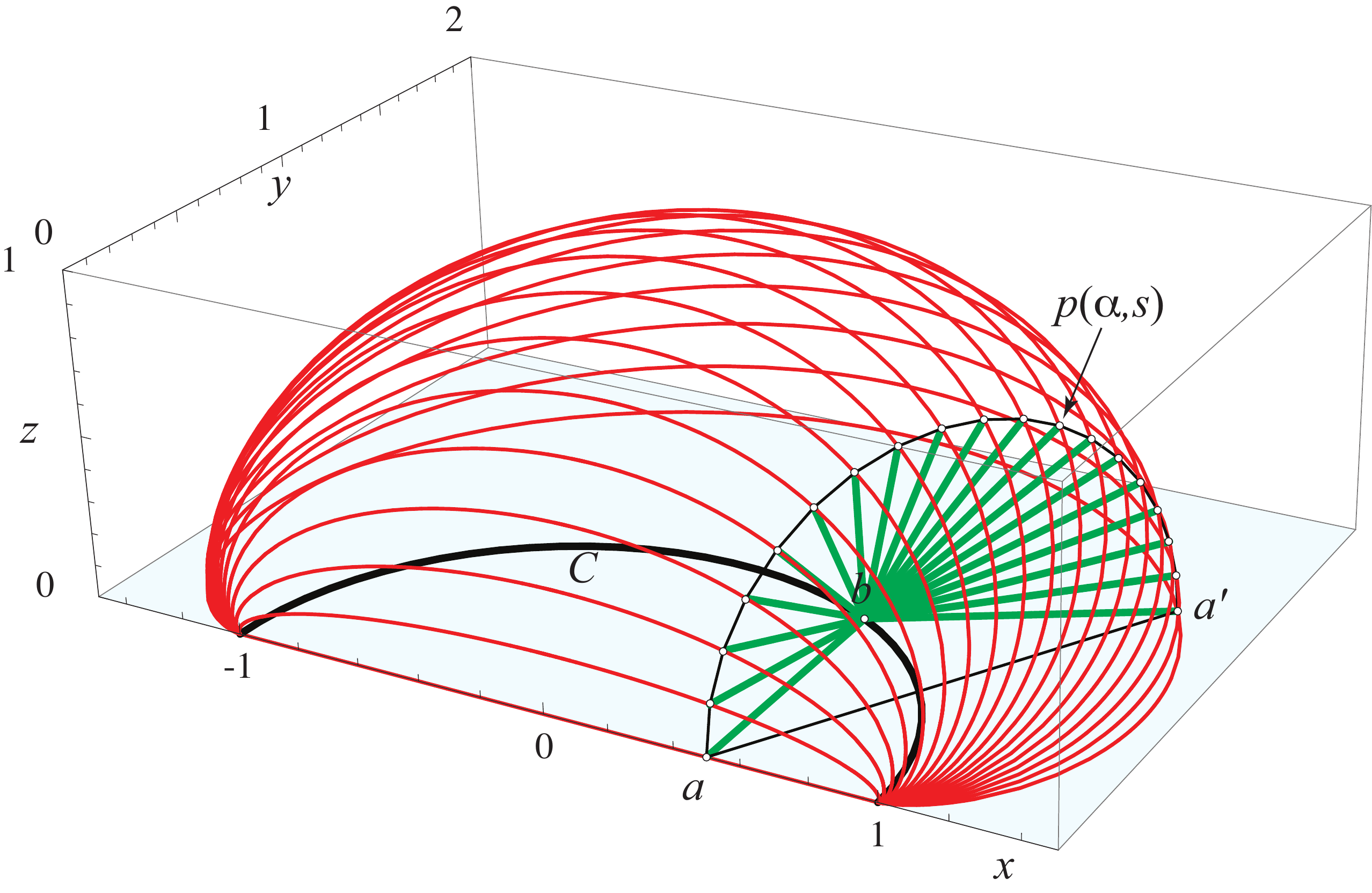}
\caption{The rib $R=ab$ sweeps out a cone as it rotates to $R'=a'b$.}
\figlab{AllCurvesCone}
\end{figure}
Recall we earlier derived the equation for the line containing $a a'$ in Eq.~\eqnref{perp.line},
which is the diameter of the semicircle traced by the rib tip.

So now we can mix three constraints to determine $p(\a,s)$:
It lies on the two spheres given by Eq.~\eqnref{spheres},
and so on their circle of intersection in the median plane $M$,
and it lies on the vertical plane $V$ through $a a'$, whose equation is Eq.~\eqnref{perp.line}.
$V$ cuts the circle on the median plane at two points, one $a$ and the other $p(\a,s)$.
See Figure~\figref{visor.3D}.
\begin{figure}[htbp]
\centering
\includegraphics[width=0.75\linewidth]{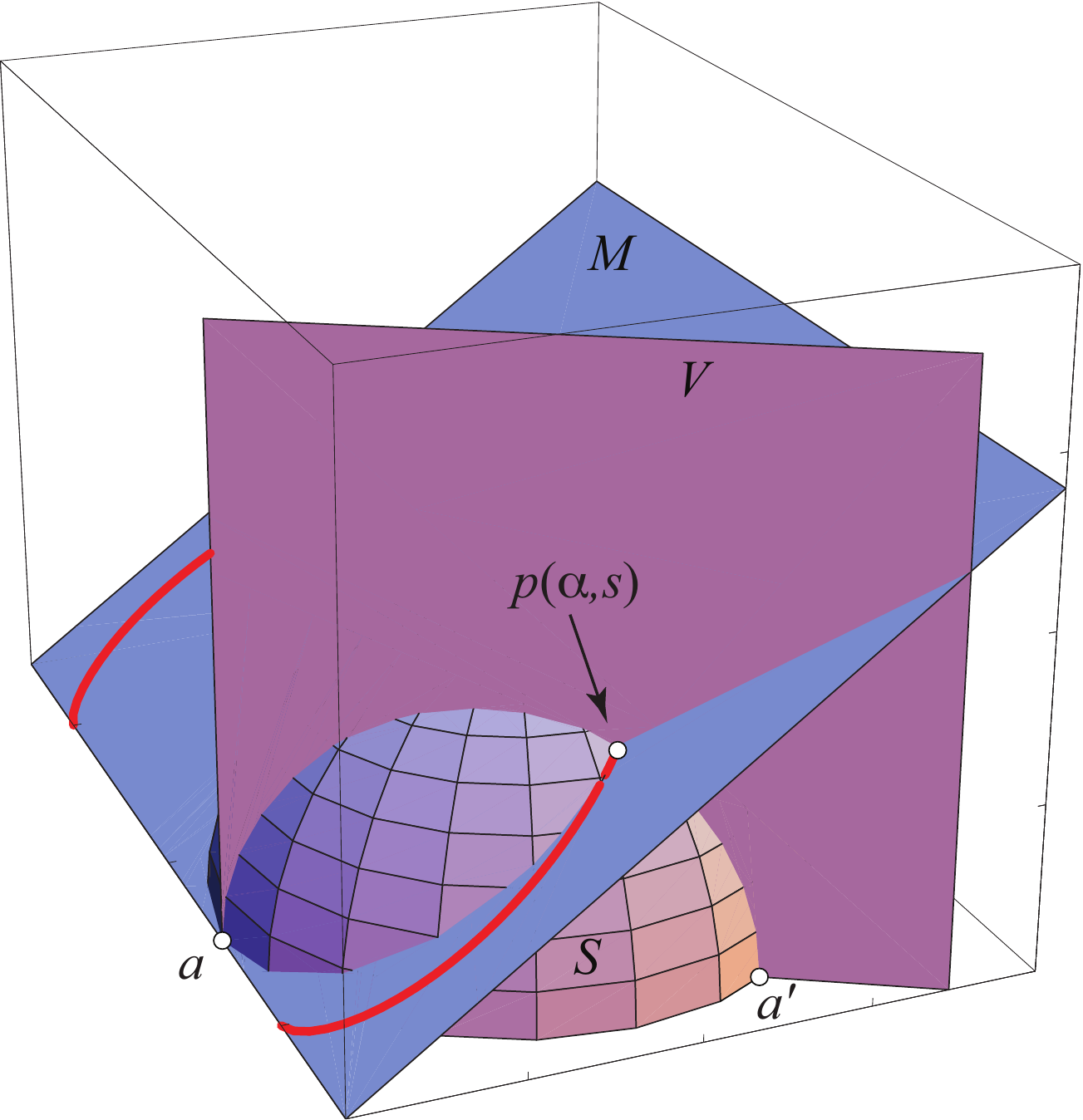}
\caption{$S \cap M$ is a circle, and $S \cap V$ is a semicircle, the rim of the cone shown in Figure~\protect\figref{AllCurvesCone}.}
\figlab{visor.3D}
\end{figure}

Solving these equations (Eqs.~\eqnref{spheres} and~\eqnref{perp.line}) simultaneously yields,
after simplification:
\begin{eqnarray*}
x(s,\a) &=&
-\frac{s
   \left(\left(s^2-2\right)
   \cos (\alpha )+3
   s^2-4\right)}{s^2 \cos
   (\alpha )-s^2+2} \nonumber \\
 y(s,\a) &=&
   \frac{4
   \left(1-s^2\right)^{3/2}
   \cos ^2\left( \tfrac{1}{2} \a \right)}{s^2 \cos
   (\alpha )-s^2+2} \eqnlab{3D} \\
 z(s,\a) &=&
   \tan \left( \tfrac{1}{2} \a \right) \;
   y(s,\a) \nonumber \;
\end{eqnarray*}
These parametric equations were used to plot
the curves in Figure~\figref{AllCurvesCone},
and in the alternate view shown in Figure~\figref{visor.curves.all}.
\begin{figure}[htbp]
\centering
\includegraphics[width=0.75\linewidth]{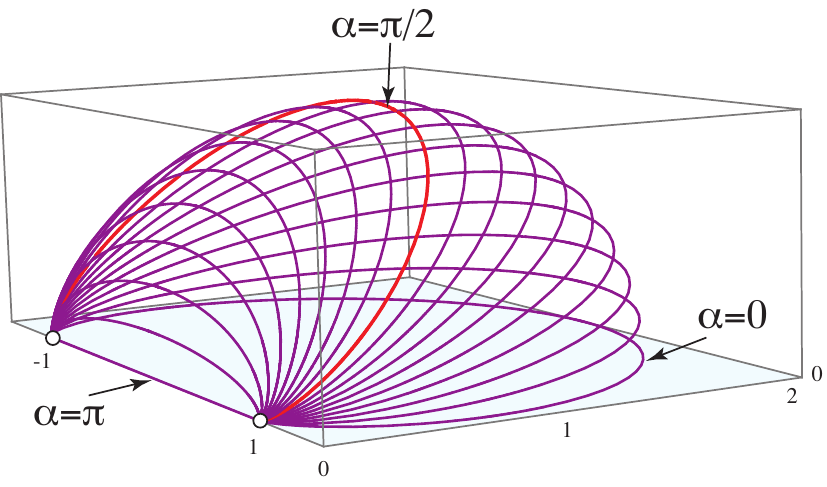}
\caption{The visor curves from $\a=\pi$ to $\a=0$.}
\figlab{visor.curves.all}
\end{figure}
Note that when $\a=\pi$, the curve reduces to the segment diameter of the circle $C$, and when $\a=0$,
the curve is the flat visor curve.
The $\a=\pi/2$ curve corresponds to the rim in the photo in Figure~\figref{knights.visor.photo}.

Although the equations Eq.~\eqnref{3D} look complicated, substitution of a fixed $\a$
of course simplifies them.
For example, with $\a=\pi/3$, they reduce to
\begin{eqnarray*}
x(s) &=&
s ( 5 -\tfrac{7}{2} s^2 ) \\
y(s) &=&
  3 (1-s^2)^{3/2} \\
z(s) &=&
   \sqrt{3}
   (1-s^2 )^{3/2} \;,
\end{eqnarray*}
which has a form recognizably similar to the parametric equations (Eq.~\eqnref{aprime}) we derived for the
flat visor curve.

%
%
%
%
%
%


\paragraph{Acknowledgments.}\label{Ack}
We thank
Molly Miller,
Duc Nuygen,
Nell O'Rourke,
Gail Parsloe,
Ana Spasova, and
Faith Weller,
who worked on pop-up cards with us in the summer of 2005.


\bibliographystyle{alpha}
\bibliography{/Users/orourke/bib/geom/geom}

\begin{thebibliography}{Wel91}

\bibitem[Jac93]{j-tpub-93}
Paul Jackson.
\newblock {\em The Pop-Up Book}.
\newblock Henry Holt and Co., 1993.

\bibitem[Wel91]{w-pdcig-91}
David Wells.
\newblock {\em The Penguin Directory of Curious and Interesting Geometry}.
\newblock Penguin, London, 1991.

\end{thebibliography}
\end{document}